\documentstyle[thmsa,sw20aip]{article}

\input{tcilatex}
\begin{document}

\author{G. H. Rawitscher$^a$, B. D. Esry$^b$, E. Tiesinga$^c$, J. P. Burke, Jr$^d$,
I. Koltracht$^e$. \\
$^a$ Physics Department, Univ. of Connecticut, Storrs, CT, 06268;\\
$^b$ITAMP,Harvard Smithonian Center for Astrophysics, Cambridge, MA 02138; \\
$^c$Atomic Physics Division, National Institute of Standards and\\
Technology,Gaithersburg, MD 20899. Permanent Address, Department of\\
Chemistry and Biochemistry, U. Maryland, College Park, MD 20742; \\
$^d$Department of Physics and JILA, Univ. of Colorado, Boulder, CO\\
80309-0440; \\
$^e$ Department of Mathematics, Univ. of Connecticut, Storrs, CT 06268}
\title{Comparison of numerical methods for the calculation of cold atom collisions.}
\maketitle

\begin{abstract}
Comparison between three different numerical techniques for solving a
coupled channel Schr\"{o}dinger equation is presented. The benchmark
equation, which describes the collision between two ultracold atoms,
consists of two channels, each containing the same diagonal Lennard-Jones
potential, one of positive and the other of negative energy. The coupling
potential is of an exponential form. The methods are i) a recently developed
spectral type integral equation method based on Chebyshev expansions, ii) a
finite element expansion, and iii) a combination of an improved Numerov
finite difference method and a Gordon method. The computing time and the
accuracy of the resulting phase shift is found to be comparable for methods
i) and ii), achieving an accuracy of ten significant figures with a double
precision calculation. Method iii) achieves seven significant figures. The
scattering length and effective range are also obtained.
\end{abstract}

\section{Introduction}

The collision between two ground state atoms at low ($\mu $K) temperature
poses challenging computational problems which can be summarized by the
words ''long range'' and ''coupled channels''. The former problem arises
from the fact that in the collision process the atomic clouds ''polarize''
each other, leading to long ranged dispersion potentials \cite{JULIENNE}.
The latter problem arises from the fact that the internal hyperfine
structure of the atoms leads to a set of coupled Schr\"odinger equations
that describe the transitions which an atom in the incident channel can make
to many of these hyperfine states. Accurate calculations of the scattering
properties and wave functions of the atom-atom collision are crucial, since
the macroscopic shape of the Bose-Einstein condensate as well as the
lineshapes of the photoassociation spectra \cite{FIORETTI} depend
sensitively on these quantities .

The lower the incident energy of the atoms, the larger are the distances for
which the potentials affect the phase shifts. This can be seen from the WKB
approximation to the phase shift which contains integrals over the local
wave length $k(x)=\left[ 2\mu (E-V(x))\right] ^{1/2}$ of the form 
\[
\int k(x)dx\simeq \int k_0dx-\frac 1{2k_0}\int V(x)dx, 
\]
since the first significant term in the expansion above contains the ratio
of the potential to the asymptotic wave number. Further, when negative
energy channels are coupled to the positive energy incident channel, it can
become difficult to enforce the appropriate decaying wave function boundary
condition, with resulting loss of stability, depending on the algorithm used.

There are various calculational methods available for dealing with this
scattering problem: modified Numerov, Gordon's \cite{GORDON}\cite{ADAM}, a
finite element method (FEM) \cite{janine}, and a recently developed method
that consist of replacing the coupled differential equations by equivalent
integral equations (IEM) \cite{IEM1}, \cite{IEM2}. In addition, there are
more sophisticated finite difference methods \cite{SIMOS}; we, however, will
not consider these methods here since they are not as widely in use. Methods
involving the representation of a continuous function by a finite set of
sampling points have been discussed \cite{KOSLOFF}. One such method led to
the mapped Fourier grid method \cite{FATTAL}, which has been employed for
the calculation of the collision between cold atoms \cite{MASNOU}. The
interaction of cold atoms with surfaces has also been discussed \cite
{BITTNER}, including how to implement boundary conditions..

Depending on the degree of accuracy required and the ease of performing the
calculation, any one of these methods may be the most suitable for a
particular situation. It is nevertheless of interest to compare these
methods with each other as far as accuracy, stability and numerical
complexity are concerned. It is the purpose of this paper to compare the
three methods, by numerically evaluating a benchmark test case described
below.

''Numerical computational stability'' is a many-faceted concept. It
manifests itself through the degree of accuracy obtained. There are at least
three ingredients:

\begin{itemize}
\item  i) How the numerical truncation error of the algorithm is offset by
the presence of the accumulation of round-off errors. The larger the number
of mesh points the smaller the truncation error, but the larger is the
corresponding overall round-off error. Different algorithms strike a
different balance between these two errors.

\item  ii)\ The sensitivity of the final result to the errors in the input
data, such as the potentials, masses, etc. This sensitivity is expresses by
the ''condition'' of the formulation model.

\item  iii) How the asymptotic boundary conditions are achieved, both in the
open and the closed channels. For the latter, the growing solutions that
contaminate the decaying solutions have to be eliminated. Each algorithm
proceeds by a different method. For example, in the IEM we introduce scaling
factors in each partition that prevent the unwanted solution. In the finite
difference methods, one has to integrate inward and outward and then match
at some intermediary distance. Here the matching matrix can introduce
errors. With the finite elements, the closed channel solutions are forced to
be zero at the final matching point, automatically eliminating the unwanted
growing solutions.
\end{itemize}

It is not the purpose of this paper to investigate in detail the stability
properties of the three algorithms described in the present study. For that,
a detailed comparison of the numerical solution with an exact solution for
an artificial test case would have to be performed. Rather, we here attempt
to obtain some numerical evidence for the degree of stability of the three
methods for a realistic example.

\smallskip\ 

\section{The Test Case}

The test case we have chosen consists of a model calculation that captures
the essence of the collision of two ultra-cold $^2$S alkali atoms. The
characteristic feature of such collisions is that in going from large to
small internuclear separations a change of coupling schemes occurs.
Asymptotically, the hyperfine structure of the individual alkali atoms
dominates, while at short internuclear separations the molecular $X^1\Sigma
_g$ and $a^3\Sigma _u$ potentials dominate. A basic understanding of the
properties of these two Born-Oppenheimer potential curves can be found in
any text book on quantum mechanics that discusses the electronic structure
of a $H_2$ molecule. Here it is sufficient to realize that they have an
identical long-range attractive van der Waals behavior and are split
exponentially via an exchange mechanism at shorter internuclear separations.

The simplest multi-channel potential that captures this physics is thus a
two channel model. Taking into account that for small collision energies the
nuclear rotation can be safely ignored, the Hamiltonian is conveniently
parametrized as 
\begin{equation}
\left\{ -\frac{\hbar ^2}{2\mu }{\bf 1}\frac{d^2}{dr^2}+\left( 
\begin{array}{cc}
V_{{\rm LJ}}(r) & Ae^{-br} \\ 
Ae^{-br} & V_{{\rm LJ}}(r)+E_{{\rm hf}}
\end{array}
\right) -{\bf 1}{\cal E}\right\} \left( 
\begin{array}{c}
\psi _P \\ 
\psi _N
\end{array}
\right) =0
\end{equation}
where the reduced mass $\mu =M/2$, $E_{{\rm hf}}$ is the asymptotic
splitting between the two channels, and ${\cal E}$ is the total energy in
the system. The $r$-dependent potentials of our test problem are the
Lennard-Jones potential $V_{{\rm LJ}}(r)=C_{12}/r^{12}-C_6/r^6$ and an
off-diagonal exchange coupling given by $Ae^{-br}$. The functions $\psi _P$
and $\psi _N$ describe the wavefunction for the open and closed channel,
respectively. Notice that the zero of energy is located at the lowest of the
two asymptotes.

Obviously this Hamiltonian is set up in terms of the atomic basis. At large
internuclear separation the Hamiltonian reduces to a diagonal matrix. In
fact, the Hamiltonian would be diagonal for all internuclear separation if
the exponential off-diagonal potential were absent. It turns out that for
internuclear separations where this term is large compared to $E_{{\rm hf}}$
it is informative to calculate the adiabatic potentials by diagonalizing the
potential term of the Hamiltonian at each internuclear separation. At
shorter distances the resulting potentials correspond to a very deep X$%
^1\Sigma _g$ and a shallow $a\Sigma _u$ potential, to a good approximation.

For two colliding ultra-cold $^2$S Na atoms, realistic values of the
constants are $M=22.9897680$ amu, $C_6$=1472 $a.u.(a_0)^6$\cite{Marinescu}, $%
C_{12}$=$38\times 10^6$ $a.u$.$(a_0)^{12}$, A=2.9 $a.u.$, b=0.81173 $a_0$,
and $E_{{\rm hf}}$= 0.2693$\cdot 10^{-6}$ a.u. This choice of $E_{{\rm hf}}$
is approximately equal to the atomic hyperfine splitting of the $^2S$ Na
atom. The total energy $E=3.1668293\times 10^{-12}a.u.$ corresponds to a
temperature of $1$ $\mu $K. Since $E\ll E_{{\rm hf}}$ , the energy in the
second channel is negative, i.e., only one of the two channels is
asymptotically accessible. In the above $a.u.$ stands for atomic units, and $%
a_0$ is the Bohr radius.

The conversion into entirely $a_0$ units is achieved by dividing the above
equation by $2\mu /\hbar ^2.$ One obtains 
\begin{equation}
\left( -\frac{d^2}{dr^2}+{\cal V-E}\right) \left( 
\begin{array}{l}
\psi _P \\ 
\psi _N
\end{array}
\right) =0,  \label{SEQ}
\end{equation}
where $r$ is in units of $a_0$ and the potential and energy matrices, ${\cal %
V\ }$and ${\cal E}$ respectively, are in units of $(a_0)^{-2}.$ The
conversion of a quantity in $a.u$. units to $(a_0)^{-2}$ units is achieved
by multiplying the former by $\mu =22.989768\times
1822.888506(a.u.)^{-1}(a_0)^{-2}.$ The potential matrix is 
\begin{equation}
{\cal V}=\left[ 
\begin{array}{ll}
\tilde V & \tilde U \\ 
\tilde U & \tilde V
\end{array}
\right] ,
\end{equation}
where $\tilde V=V\times \mu $, $U=\tilde U\times \mu $, and the energy
matrix is 
\begin{equation}
{\cal E}=\left[ 
\begin{array}{ll}
k^2 &  \\ 
& -\kappa ^2
\end{array}
\right] .
\end{equation}
Here $k$ and $\kappa $, the wave numbers in each channel, are given by $k=%
\sqrt{E\times \mu }$ and $\kappa =\sqrt{E_{hf}\times \mu -k^2}.$ In our
numerical example, the corresponding values are $k=3.643004224146145\times
10^{-4}(a_0)^{-1}$\ and $\kappa =0.1062338621818394(a_0)^{-1}.$ The wave
function is normalized so that asymptotically it becomes 
\begin{equation}
\left( 
\begin{array}{l}
\psi _P \\ 
\psi _N
\end{array}
\right) \approx \left( 
\begin{array}{l}
\sin (kr)+K_1\cos (kr) \\ 
K_2\exp (-\kappa r)
\end{array}
\right) ,  \label{DEF}
\end{equation}
where $K_1$ and $K_2$ are two elements related to the real scattering $R$
matrix, in terms of which the phase shifts can be obtained.

In this model, the diagonal potential extends to considerably larger
distances than the coupling potential. Further, between 6 and 10 $a_0$ the
diagonal potential is very deep leading to many oscillations in the wave
functions. For example, near $5.5a_0$ the local wave length $\lambda $ in
both channels is $\simeq 0.25a_0,$ near $8.5a_0$ $\lambda \simeq 1.2a_0,$
with smaller ripples superimposed, and near $20a_0$ the local wave length
has increased to $\simeq 4a_0$ At distances less than $4a_0$ the repulsive
portion of the potential becomes very large making the wave function very
small. In order to allow for the singularity of the diagonal potentials near
the origin, a parameter $R_{cut}$ is defined, and the wave functions are set
to zero in the interval $[0,R_{cut}].$ A value of $R_{cut}=4.0a_0$ is found
to be satisfactory. In addition, the calculation is carried out to a maximum
radius, $R_{\max }$, beyond which all potentials are set equal to zero. In
our calculations, $R_{\max }$ is set equal to $500a_o.$ When $R_{\max }$ is
increased further, the values of $K_1$ and $K_2$ still change beyond the 6
th. significant figure, as can be seen from the Table in Appendix, even
though the Lennard-Jones potential is less than $-3.95\times 10^{-9}a_0^{-2}$%
. The large effect on the phase-shift produced by such a small potential is
due to the occurrence of the factor $1/k\simeq 2.75\times 10^3$ in the
integrals involving the potential tail, as was already pointed out in the
introduction. Rather than numerically calculating such changes, it is
preferable to employ perturbation methods, which are described in Appendix 2.

\subsection{The Integral Equation Method.}

In this method the differential equation $\left( \frac{d^{2}}{dr^{2}}+{\cal E%
}\right) \psi ={\cal V}\psi $ is transformed into the Lippmann-Schwinger
integral equation 
\[
\psi (r)=F(r)+\int_{0}^{R_{\max }}{\cal G}_{0}(r,r^{\prime }){\cal V(}%
r^{\prime })dr^{\prime }, 
\]
where $F(r)$ is a undisorted wave function, like $\left( \sin (kr),0\right)
, $ and ${\cal G}_{0}(r,r^{\prime })$ is the undistorted Green's function
matrix \cite{IEM2}. A motivation for such an approach is that the solutions
of integral equations have better numerical stability than the solutions of
differential equations. One common objection to the use of integral
equations has been that the solution leads to large matrices which are not
sparse and hence require substantially larger amounts of computing time than
the sparse matrices of differential equations. This objection was overcome
in our integral equation method (IEM) by dividing the whole radial interval
into partitions. The integral equations in each partition lead to dense
matrices of small dimension, but the matrix that combines the local
solutions into the global one, albeit of large dimension, is sparse. It
should be noted that this latter property is valid only in configuration
space, because only in this space do the Green's functions have the required
semi-separable nature. In the present version of the IEM method the
(variable) size of each of the partitions is determined in terms of two
parameters $NL$ and $\epsilon $ as follows. In each radial region a local
wave length in channels 1 and 2 is obtained as $2\pi /\sqrt{|k^{2}-\tilde{V}%
(r)|},$ and $2\pi /\sqrt{|-\kappa ^{2}-\tilde{V}(r)|}.$ The smaller of the
two local wavelengths is taken, and the size of the partition in that region
is determined such that there are a given total number $NL$ of Chebyshev
points per local wave length. Allowing for the fact that in each partition
there are 16 Chebyshev points, the average length of a partition for a given
local wave length $\lambda $ is $\lambda \times 16/NL.$ The length of each
partition is subsequently readjusted using the tolerance parameter $\epsilon 
$ as follows. According to the IEM method \cite{IEM1}, in each partition two
sets of ''local'' functions are calculated in terms of which the global
function $\psi $ is obtained as a linear combination. The accuracy of each
of the local functions can be determined by the size of the coefficients of
the highest order Chebyshev polynomials. If the relative accuracy of the
local functions in a given partition is larger than $\epsilon ,$ then that
partition is divided in half, and the testing is continued. If the initially
chosen value of $NL$ is too small, then the initial partitions are too
large, and many of the partitions are subsequently reduced by the $\epsilon $
criterion. In this case the final number of partitions $M$ becomes larger
than their initial value. If the chosen value of $NL$ is too large, then
most of the partitions are unnecessarily small, and the value of $M$ is too
large, leading to a larger accumulation of roundoff errors for the final
elements of the $K$-matrix. An exception is the interval $[0,R_{cut}].$ This
interval is considered as one partition, containing a total of $16$
Chebyshev points. This is sufficient since the wave function is very small
in this region (less than the desired accuracy for the values of $K$ ), and
the values of $K$ were found to be stable to 11 significant figures as $%
R_{cut}$ was varied below $4.0a_{0}$ .

In summary, for a given value of $\epsilon ,$ the value of $NL$ was varied
until the smallest number of partitions $M$ was obtained. An example is
given in the table below.\medskip \medskip

\begin{center}
Table 1. Values of $K_1$ and number of partitions $M$ as a\\[0pt]
function of NL for the tolerance $\epsilon .=10^{-9}$\medskip \medskip

\begin{tabular}{|cccc|}
\hline
\multicolumn{1}{|c|}{$NL$} & \multicolumn{1}{c|}{$M$} & \multicolumn{1}{c|}{$%
K_1$} & $K_2$ \\ \hline
\multicolumn{1}{|c|}{10} & \multicolumn{1}{c|}{150} & \multicolumn{1}{c|}{
-0.31233398338809} & 6.5761303971514 \\ \hline
\multicolumn{1}{|c|}{20} & \multicolumn{1}{c|}{153} & \multicolumn{1}{c|}{
-0.31233398339572} & 6.5761303973071 \\ \hline
\multicolumn{1}{|c|}{30} & \multicolumn{1}{c|}{144} & \multicolumn{1}{c|}{
-0.31233398339229} & 6.5761303972290 \\ \hline
\multicolumn{1}{|c|}{40} & \multicolumn{1}{c|}{154} & \multicolumn{1}{c|}{
-0.31233398339070} & 6.5761303972039 \\ \hline
\multicolumn{1}{|c|}{50} & \multicolumn{1}{c|}{177} & \multicolumn{1}{c|}{
-0.31233398338870} & 6.5761303971639 \\ \hline
\end{tabular}
\end{center}

\medskip \medskip From this table one can find a value of $K_1=-0.3123339834$
and $K_2=6.576130397$ which are stable to ten significant figures. For
values of the tolerance $\epsilon $ between $10^{-13}$ and $10^{-3}$ a good
compromise value of $10$ for $NL$ was found. The corresponding values of $M$
and the corresponding accuracy of $K_1$ are listed in Table 2 for several
values of $\epsilon .$ \medskip \medskip

\begin{center}
Table 2. Accuracy for $K_1$ and number of partitions $M$ for a given\\[0pt]
value of the Tolerance $\epsilon $, with $NL=10$\medskip

\begin{tabular}{|c|c|c|c|}
\hline
$\epsilon $ & $M$ & $K_1$ & \# of Sign. Figs. \\ \hline
$10^{-2}$ & 28 & {\bf -0.312}43337402099 & 3 \\ \hline
$10^{-3}$ & 30 & {\bf -0.31233}467247746 & 5 \\ \hline
$10^{-5}$ & 67 & {\bf -0.31233398}457315 & 8 \\ \hline
$10^{-7}$ & 106 & {\bf -0.31233398}370637 & 8 \\ \hline
$10^{-9}$ & 158 & {\bf -0.3123339833}8809 & 10 \\ \hline
$10^{-11}$ & 214 & {\bf -0.3123339833}8869 & 10 \\ \hline
$10^{-13}$ & 574 & {\bf -0.3123339833}8534 & 10 \\ \hline
\end{tabular}
\medskip \medskip
\end{center}

From this table it appears that beyond $\epsilon =10^{-9}$ the accumulation
of roundoff errors begins to dominate, and -0.31233398339 is the best value
of $K_{1}.$ The distribution of partitions for three tolerance parameters is
shown in Fig. 1.\FRAME{ftbphFU}{5.3999in}{3.4497in}{0pt}{\Qcb{Various IEM
partition distributions, described in Table 2. The y-axis represents the
partition number $i,$ and the x-axis shows the lower boundary of partition $%
i.$ The more points in a particular radial interval, the smaller are the
lenghts of the partitions in that interval. All three partition
distributions started with the same number of mesh points per local
wavelength (NL = 20), but were subsequently modified by the $\epsilon $%
-accuracy criterion, with $\epsilon =10^{-9},10^{-6},$ and $10^{-3},$
respectively. The numbers above each curve represent the number of accurate
significant figures achieved for the asymptotic constant $K_{1}$ for each
value of $\epsilon .$ The large concentration of partitions in the vicinity
of $50a_{0}$ reflects the occurrence of a turning point in the negative
energy channel near that distance, where accuracy would have been lost had
the initial partition distribution been used.}}{}{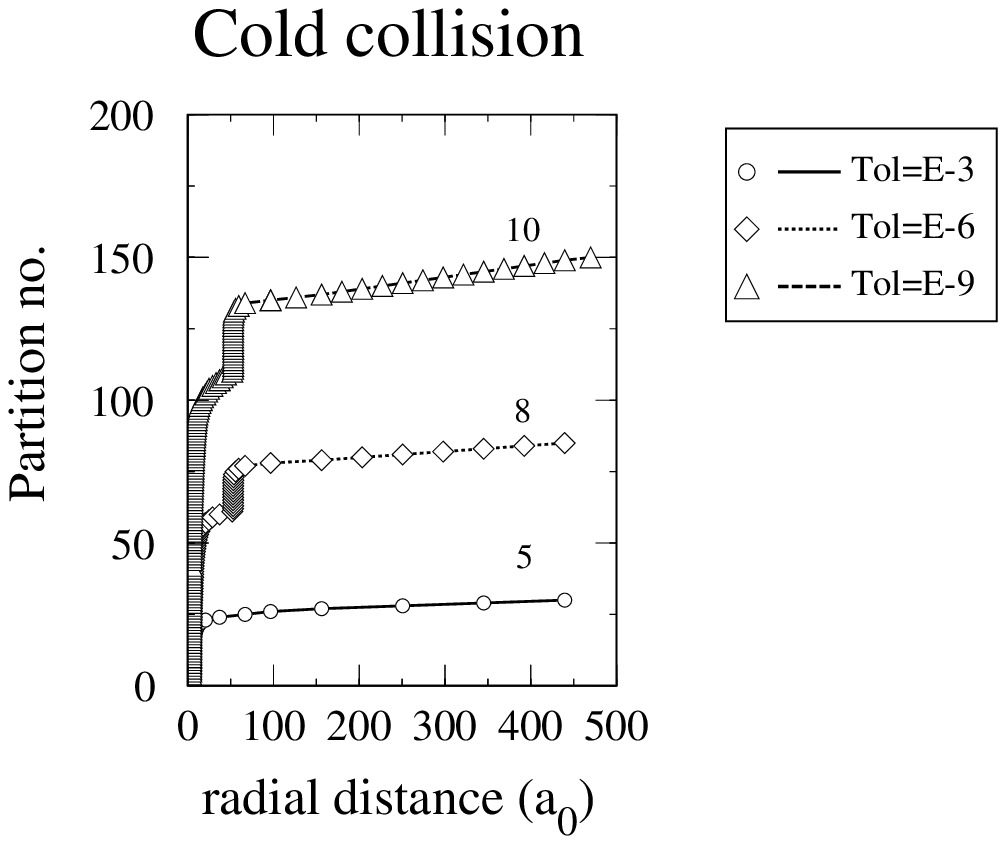}{%
\special{language "Scientific Word";type "GRAPHIC";maintain-aspect-ratio
TRUE;display "USEDEF";valid_file "F";width 5.3999in;height 3.4497in;depth
0pt;original-width 388.4375pt;original-height 247.9375pt;cropleft
"0";croptop "1";cropright "1";cropbottom "0";filename
'A:/iemeitop.eps';file-properties "XNPEU";}}The increasingly large spacing
of the partitions at the large distances is clear from the figure. In the
vicinity of $R\simeq 50$ the density of partitions is high because the
negative energy channel has a turning point there. This shows that the local
wave-length criterion alone would have been insufficient to determine the
partition size.

\subsubsection{Scattering Length and Effective Range.}

As a further test of the stability of the IEM method, the Scattering Length $%
a$ and Effective Range $r_{e}$ are investigated. They are obtained in the
limit of small wave number k from the expression 
\begin{equation}
k/K_{1}=-\frac{1}{a}+r_{e}k^{2}+O(k^{3}).  \label{scattl}
\end{equation}
The left hand side of the above equation is usually written as $k\cot \delta
_{0},$ which is equal to $k/K_{1}.$ For a given choice of the truncation
radius $R_{\max }$ the value of $K_{1}$ is calculated for two different and
small values of $k$ and the values of $a$ and $r_{e}$ are obtained from the
two values of $k/K_{1}$ in Eq. (\ref{scattl}).

However, as mentioned in the introduction, the values of the scattering
matrix $K$ depend on the choice of the truncation radius, the more so the
smaller the value of $k$ , because of the increasingly non-negligible
contributions of the potential beyond $R_{\max }.$ Hence, in order to obtain
a reliable value of the scattering length, it is advantageous to first
correct $K_{1}$ for the contributions beyond $R_{\max }.$ This can be done
quite simply by using first order perturbation theory, as is detailed in
Appendix 1, and as will be demonstrated below.

An example of the variation of the $K$'s with $R_{\max }$, obtained by the
IEM, is given in the table below. All results in this section were obtained
with NL = 10, and $\epsilon =10^{-9}.$ \medskip .

\begin{center}
Table 3: Dependence of $K_1$ and $K_2$ on $R_{\max }.$\medskip \\[0pt]
\medskip

$
\begin{tabular}{|c|c|c|c|}
\hline
$K_{1}(R_{\max })$ & $K_{1}(\infty )$ & $K_{2}(R_{\max })$ & $R_{\max }$ \\ 
\hline
-.313705209 & -.312322902025 & 6.62005410 & 250 \\ \hline
-.312333983 & -.312323344009 & 6.57613040 & 500 \\ \hline
-.312324588 & -.312323343934 & 6.57555968 & 1000 \\ \hline
-.312324073 & -.312323343936 & 6.57558157 & 1500 \\ \hline
-.312323719 & -.312323343936 & 6.57558741 & 2000 \\ \hline
\end{tabular}
$
\end{center}

\medskip One can see from the table that the value of $K_1(R_{\max })$
becomes monotonically less negative as $R_{\max }$ increases, while $K_2$
first decreases, and then increases for $R_{\max }>1000.$ This behavior can
be reproduced numerically by means of perturbation theory, described in the
appendix. The usefulness of the perturbative correction is also demonstrated
by the stability of the column denoted as $K_1(\infty ),$ which contains the
corrected values of $K_1(R_{\max }).$ The table shows that the perturbative
correction increases the stability of $K_1$ from 5 or 6 significant figures
to 11, yielding $K_1(\infty )=-.31232334394$

The stability of the values of $a$ and $r_e$ will be described next. The two
smallest values of $k$ to be used in Eq. (\ref{scattl}) were approximately $%
0.1152\times 10^{-4}$ and $0.3643\times 10^{-6},$ for which the third order
term is smaller than the accuracy of the present method. Their values with
and without the perturbative correction of $K_1$ are shown in the table
below.

\medskip .

\begin{center}
Table 4: Dependence of $a$ and $r_e$ on $R_{\max }.$\medskip \medskip

$
\begin{tabular}{|c|c|c|c|c|}
\hline
$a(R_{\max })$ & $a(\infty )$ & $r_e(R_{\max })$ & $r_e(\infty )$ & $R_{\max
}$ \\ \hline
852.0123407 & 851.9817159134 & 55.08319944 & 55.1051720100 & 500 \\ \hline
851.9849554 & 851.9817157297 & 55.10378197 & 55.1051694472 & 1000 \\ \hline
851.9837969 & 851.9817157354 & 55.10138602 & 55.1051696827 & 1500 \\ \hline
851.9829235 & 851.9817157362 & 55.10046968 & 55.1051693964 & 2000 \\ \hline
851.9824507 & 851.9817157362 & 55.10031183 & 55.1051691577 & 2500 \\ \hline
\end{tabular}
$
\end{center}

\medskip

The table shows that the stability of the scattering length is increased
from 5 to 11 significant figures by the perturbative correction, yielding $%
a(\infty )=851.98171574.$ Likewise, the stability of the effective range is
increased from 5 significant figures to 8, yielding $r_e(\infty )=55.105169.$

\subsection{The Finite Element method}

The second method employs the non-iterative eigenchannel \cite{chg} variant
of the R-matrix method first introduced by Wigner and Eisenbud \cite{Wigner}%
. The eigenchannel R-matrix method solves the Schr\"{o}dinger equation
within a finite reaction volume $\Omega $ of configuration space, subject to
constant normal logarithmic derivative boundary conditions on the surface $%
\Sigma $ of $\Omega $. The collisional properties of the system, typically
represented in terms of an $S$-matrix, are easily obtained once the normal
logarithmic derivative $b=(\partial \Psi /vitial n)\Psi ^{-1}$ is calculated.

One can obtain the following eigenvalue equation for the normal logarithmic
derivative $b$ on the surface $\Sigma $\cite{chg}: 
\begin{equation}
\underline{\Gamma }\vec{c}=b\underline{\Lambda }\vec{c}  \label{eigen}
\end{equation}
We solve this equation using the finite-element method (FEM) \cite{janine}.
The FEM divides the radial domain into $N$ sectors (or partitions) and
within each sector defines a local basis in much the same spirit as the IEM
above. The local basis functions, however, are fifth-order Hermite
interpolating polynomials rather than Chebyshev polynomials. The Hermite
interpolating polynomials $u_{k}(x_{n})$, $k$=1--6, are non-zero only in
sector $n$. Here, $x_{n}$ is a rescaled variable defined on the interval $%
\left[ -1,1\right] $ that is related to the physical internuclear separation 
$R$ in sector $n$ through an appropriate linear transformation. The six
basis functions are defined through the following boundary conditions: 
\begin{equation}
\begin{array}{lll}
u_{k}(-1)=\delta _{1k} & u_{k}(0)=\delta _{3k} & u_{k}(1)=\delta _{5k} \\ 
u_{k}^{\prime }(-1)=\delta _{2k} & u_{k}^{\prime }(0)=\delta _{4k} & 
u_{k}^{\prime }(1)=\delta _{6k}
\end{array}
\;.
\end{equation}
A multi-component radial wave function can then be represented by the
following expansion on the piecewise polynomials $u_{i}(x)$: 
\begin{equation}
\Psi (R)=\sum_{i\equiv \{m,k,n\}}c_{i}u_{i}(x_{n})\;.
\end{equation}
The set $i$ contains the basis function index $k$, the channel index $m$,
and the sector index $n$. The coefficients $c_{i}$ are to be determined
subject to a continuity constraint on each channel component and its first
derivative across sector boundaries. The assembly stage of the calculation
is thus performed prior to the main calculation, in contrast to the IEM
where it is performed after. In addition, channel boundary conditions can be
imposed quite simply by setting the value of the appropriate coefficient.
For instance, the closed channel function is forced to be zero on the
surface by setting the coefficient of $u_{5}$\ in the last sector to zero.

In the finite element representation, the matrix elements of $\underline{%
\Gamma }$ and $\underline{\Lambda }$ in Eq.(\ref{eigen}) are given by 
\begin{equation}
\Gamma _{ij}=2\mu \int_{-1}^1u_i(x_n)(E-H)u_j \\
(x_n)a_ndx_n-\delta _{m,m^{\prime }}\delta _{n,N}\delta _{k,5}\delta
_{k^{\prime },6}/a_n
\end{equation}
and 
\begin{equation}
\Lambda _{ij}=\delta _{m,m^{\prime }}\delta _{n,N}\delta _{k,5}\delta
_{k^{\prime },5}\,,
\end{equation}
respectively, where $a_n=(R_{n+1}-R_n)/2$. Here $i\equiv \{m,k,n\}$ and $%
j\equiv \{m^{\prime },k^{\prime },n\}$, and $H$ is the Hamiltonian of the
system. Because the basis functions are non-zero only within a given sector,
the corresponding matrices have roughly a block diagonal structure. Each
sector has an associated block which is coupled only to its nearest
neighbors through the continuity conditions. Note that the overlap matrix $%
\underline{\Lambda }$, whose elements are given by surface integrals over
the basis functions\cite{chg}, is particularly simple in the FEM
representation. It contains only $m_o$ non-zero elements (all equal to one),
where $m_o$ is simply the number of channels open (i.e., $E>V_m$) on the
surface $\Sigma $. The $\underline{\Gamma }$ matrix is symmetric and, as
mentioned, can be constructed in a banded format. The integrals representing
the matrix elements of $\underline{\Gamma }$ are also particularly simple in
the FEM representation. In fact, except for the integral over the
interaction potential, all integrals can be evaluated analytically once and
for all before hand, significantly decreasing the CPU time necessary to
construct the matrix.

At this stage, we are left with a banded generalized eigensystem to solve
that typically has large dimensions but has at most only $m_o$ non-zero
eigenvalues. Although there are standard linear algebra packages which could
solve these equations directly, implementing an efficient, general method
would be difficult since the non-zero eigenvalues can range between $-\infty 
$ and $+\infty $ and all $m_0$ of them are needed. It has been shown \cite
{chg}, however, that by partitioning the matrices according to whether the
basis functions are non-zero (open $\equiv o$) or zero (closed $\equiv c$)
on $\Sigma $, i.e. by writing 
\begin{equation}
\left( 
\begin{array}{cc}
\underline{\Gamma }^{cc} & \underline{\Gamma }^{co} \\ 
\underline{\Gamma }^{oc} & \underline{\Gamma }^{oo}
\end{array}
\right) \left( 
\begin{array}{c}
c^c \\ 
c^o
\end{array}
\right) =b\left( 
\begin{array}{cc}
0 & 0 \\ 
0 & \underline{\Lambda }^{oo}
\end{array}
\right) \left( 
\begin{array}{c}
c^c \\ 
c^o
\end{array}
\right) ,
\end{equation}
Eq. (\ref{eigen}) can be reduced to a small ($m_o\times m_o$) eigensystem 
\begin{equation}
\underline{\Omega }^{oo}\,\vec {c^o}=b\,\underline{\Lambda }^{oo}\,\vec {c^o}
\label{eigen2}
\end{equation}
where $\underline{\Omega }^{oo}=\underline{\Gamma }^{oo}-\underline{\Gamma }%
^{oc}(\underline{\Gamma }^{cc})^{-1}\underline{\Gamma }^{co}$. The main
computational burden is thus shifted to constructing $\underline{\Omega }%
^{oo}$. Since $\underline{\Gamma }^{cc}$ is a large banded matrix in the FEM
representation, this is most efficiently accomplished by solving the set of
linear equations 
\begin{equation}
\underline{\Gamma }^{cc}\vec X=\underline{\Gamma }^{co}.  \label{eigen3}
\end{equation}

The solution $\vec X=(\underline{\Gamma }^{cc})^{-1}\underline{\Gamma }^{co}$
thus provides the needed matrix inverse. $\underline{\Gamma }^{cc}$ has
dimensions $4MN$$\times $$4MN$, but with a half-bandwidth (number of
non-zero diagonals above the main diagonal) of only $6M-1$ resulting from
the FEM representation. $\underline{\Gamma }^{co}$ has dimensions $4MN$$%
\times $$m_o$. We use standard LAPACK\cite{Lapack} routines to solve
equations \ref{eigen2} and \ref{eigen3}. The eigenvalues $b$ and
eigenvectors $\vec c^o$ completely specify the linearly independent
solutions of the Schr\"odinger equation on the surface $\Sigma $. This
information is generally packaged in terms of a $R$-matrix 
\begin{equation}
R_{mm^{\prime }}=\sum_\beta Z_{m\beta }b_\beta ^{-1}Z_{\beta m^{\prime
}}^{-1}  \label{R}
\end{equation}
where the columns of $Z$ are given by the eigenvectors $\vec c^o$. The $S$%
-matrix is then obtained through simple matrix manipulations involving only
the $R$-matrix and the two linearly independent solutions of the asymptotic
form of the Schr\"odinger equation\cite{chg}.

For the two-channel test problem described above, Eq. (\ref{eigen3}) is a $%
8N $$\times $$8N$ matrix equation with a half-bandwidth of 11, and only one
solution is required since there is only a single open channel, $m_{o}$=1.
The present formulation of the R-matrix method focuses the computational
effort on finding only the relevant scattering information. Thus, only the
coefficient in the open channel $K_{1}$ is obtained. The results are shown
in the table below as a function of the number of sectors $N$ used.\medskip
\medskip

\begin{center}
Table 5. Results for the Finite Element Method\medskip \medskip

\begin{tabular}{|c|c|}
\hline
$K_1$ & $N$ \\ \hline\hline
$-0.312345739663914$ & 400 \\ 
$-0.312334009008278$ & 800 \\ 
$-0.312334008856244$ & 1600 \\ 
$-0.312334008759115$ & 3200 \\ 
$-0.312334006921697$ & 6400 \\ \hline
\end{tabular}

\medskip \medskip
\end{center}

\subsection{The Gordon Algorithm.}

The Gordon algorithm \cite{GORDON}, \cite{ADAM} is a well-established
numerical method to solve for the scattering solutions of a set of $N_c$
coupled radial Schr\"odinger equations. Similar to the FEM discussed in the
previous section, this algorithm is local in the sense that the wavefunction
is propagated from $R$ to $R+dR$ using only the wavefunction at $R$. The
version of the Gordon method used here is described in Ref. \cite{MIES}. The
main difference from the original Gordon method is in replacing the Airy
functions, which correspond to the reference solutions for linearized
potentials, by $\sin qr$ or $\cos qr$ in each interval $dR.$ Here the
quantities $q$ are the local wavenumbers of each channel in the interval $dR$
. They are obtained by replacing the potential matrix in interval $dR$ by an
averaged constant potential, and then diagonalizing this potential.{\bf \ }%
If the local energy is negative then $\exp (qr)$ or $\exp (-qr)$ are used
instead. The wave function in the diagonalized interval $dR$ is represented
in each channel $n$ by $\Psi _n=A_n(r)\sin q_nr+B_n(r)\cos q_nr,$and the
vector of the coefficients $A$ and $B$ are calculated by solving a system of
first order coupled differential equations involving the difference between
the true and the averaged potential matrix in interval $dR$. In the
intervals where some of the local energies are negative (i.e., some of the
q's are imaginary), the undesired exponentially increasing functions are
minimized by a ''triangularization'' method developed by Gordon. Analytic
connection formulas between $R$ and $R+dR$ determine $A$ and $B,$ and thus
the solution is propagated towards the final matching point. Consequently no
large linear system needs to be evaluated in the Gordon algorithm and hence
the method is not memory limited.

The physical boundary conditions for the solution are obtained by first
calculating N$_c$ linear independent solutions from different initial
conditions near the origin, and then constructing appropriate linear
combination of these ''mathematical'' solutions. The stepsize $dR$ is
calculated on the fly. After a fixed number of steps (say ten) the
percentage change of the $A$ and $B$ coefficients between intervals $n$ and $%
n+i$, where $i\simeq 10$, is compared to an accuracy criterium. If the
variation is too large, the step size is repeatedly halved until the
accuracy criterion is met. On the other hand if the coefficients hardly
changed the step size is doubled. Consequently, if the potential is well
approximated by a constant, large steps are taken. Closed channels are
removed from the propagation when the corresponding amplitude of the channel
wavefunction has become smaller than a threshold parameter. The influence of
these components on the scattering properties of the asymptotically
accessible channel is then negligible.

Our results with this algorithm for $R_{cut}$=4 $a_0$ and $R_{max}$=500 $a_0$
are summarized in Table 6. The columns describe the fractional change of the 
$A$ and $B$ coefficients, the number of steps, and $K_1$, respectively. It
is immediately clear that the Gordon method uses many more steps than the
other two numerical methods. As discussed before this is not crucial as it
is not necessary to store the wavefunction at every step in order to
propagate the wavefunction. \medskip

\begin{center}
Table 6. Results for the Gordon Method\medskip \medskip\medskip

\begin{tabular}{|l|l|l|}
\hline
Fract'l & No of &  \\ \hline
Change & steps & $K_{1}$ \\ \hline
0.1000 & 7025 & -0.312348088 \\ \hline
0.0500 & 7632 & -0.312340564 \\ \hline
0.0250 & 8353 & -0.312337479 \\ \hline
0.0100 & 9796 & -0.312335340 \\ \hline
0.0050 & 11402 & -0.312334854 \\ \hline
0.0020 & 14454 & -0.312334070 \\ \hline
0.0010 & 18179 & -0.312334067 \\ \hline
0.0005 & 25502 & -0.312334020 \\ \hline
\end{tabular}
\medskip
\end{center}

\section{Discussion}

A comparison between the three methods is displayed in Table 7, which lists
the number of mesh points required to achieve a certain accuracy for $K_1.$

\medskip \medskip

\begin{center}
Table 7. Comparison between three methods.\medskip \medskip

$
\begin{tabular}{|l|l|l|l|l|}
\hline
&  & No of & No of & Total No \\ \hline
& $K_1$ & partit's & Pts/part. & \multicolumn{1}{|l}{of Points..} \\ 
\hline\hline
IEM & -0.3123339834 & 153 & 16 & 2,448 \\ \hline
FEM & -0.312334009 & 800 & 4 & 3,200 \\ \hline
Gordon & -0.3123340 & 25,502 & 1 & 25,502 \\ \hline
\end{tabular}
$\medskip
\end{center}

The finite element (FEM) and the integral equation (IEM) methods are nearly
identical in performance. Both can easily adjust the size of the partitions
to the local conditions of the potentials; both give stability of at least
nine significant figures, and both use approximately the same number of mesh
points. Their numerical complexity is also comparable since a CPU-time test
shows that both use approximately the same computer time for the case tested
here. The IEM\ and the FEM differ in the 8th significant figure (by $%
2.6\times 10^{-8})$ for $K_{1}.$ The reason for this difference is not
known, but could be related to the fact that both the FEM and the Gordon
methods have not yet fully converged as the number of partitions is
increased, as can be seen from Tables 5 and 6, and in Figs. 2 and 3. \FRAME{%
ftbpFU}{5.399in}{3.4549in}{0pt}{\Qcb{Comparison of the rate of convergence
of the numerical value of $K_{1}$ as a function of the total number of
mesh-points N. On the vertical axis is plotted the value of $K_{1}$ from the
fifth significant figure onwards. This is accomplished by subtracting
-0.312300000 from each value of $K_{1}$ and multiplying the result by $%
10^{4}.$ The values of $K_{1}$ are taken from Tables 2, 5, and 6 for the
IEM, FEM and Gordon methods. They are represented by solid circles, open
circles and squares, respectively}}{}{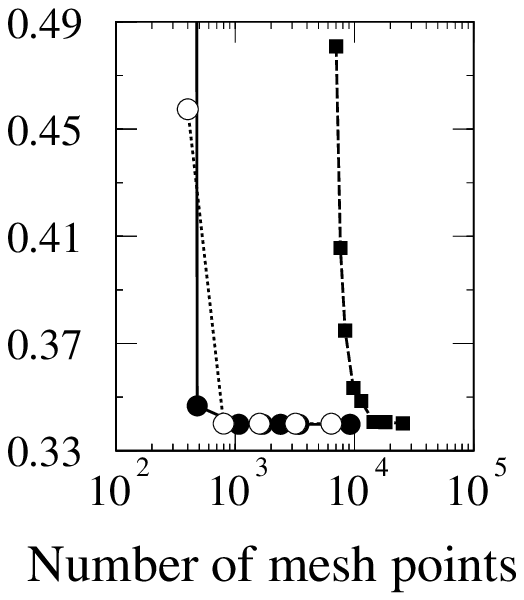}{\special{language
"Scientific Word";type "GRAPHIC";maintain-aspect-ratio TRUE;display
"USEDEF";valid_file "F";width 5.399in;height 3.4549in;depth
0pt;original-width 388.4375pt;original-height 247.9375pt;cropleft
"0";croptop "1";cropright "1";cropbottom "0";filename
'A:/coc2.eps';file-properties "XNPEU";}}\FRAME{ftbphFU}{5.399in}{3.4549in}{%
0pt}{\Qcb{Same as Fig. 2 with a larger magnification.}}{}{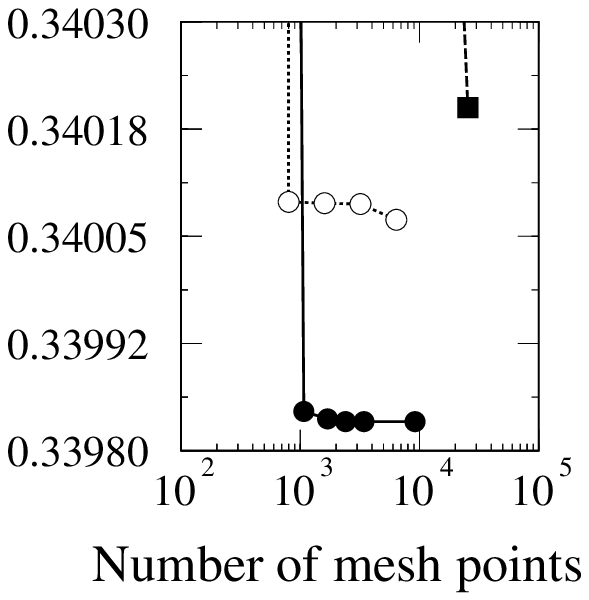}{%
\special{language "Scientific Word";type "GRAPHIC";maintain-aspect-ratio
TRUE;display "USEDEF";valid_file "F";width 5.399in;height 3.4549in;depth
0pt;original-width 388.4375pt;original-height 247.9375pt;cropleft
"0";croptop "1";cropright "1";cropbottom "0";filename
'A:/coc1.eps';file-properties "XNPEU";}}These figures show the value of $%
K_{1},$ from the fifth significant figure onward, as a function of the total
number of mesh points for each of the three methods in progressively larger
detail. The somewhat slower convergence for the FEM may be due to the fact
that the algorithm for determining the partition size, especially in the
vicinity of a turning point, is not as refined for the FEM as for the IEM.

The Gordon method is simpler to implement than the two other ones and gives
a respectable accuracy of seven significant figures, albeit at the expense
of a much larger number of mesh points, which in turn leads to a larger
accumulation of roundoff errors. An improved Numerov method gave only four
significant figures ($K_1$= -0.31233(2)) and is not mentioned further. In
the IEM, the boundary conditions are built in automatically via the Green's
functions, while in the present form of the FEM the solutions with
exponential growth are eliminated by forcing the closed channel component to
be zero on the surface $\Sigma .$ In a separate study of two coupled
equations \cite{IEM2}, it is shown that there are situations in which the
conventional Numerov method has severe difficulty in obtaining the correct
asymptotic boundary condition, while both the IEM and the FEM do not.

It should be clear that the accuracy achieved in this benchmark calculation
is not directly relevant for physical applications, since the potentials are
usually known only to low accuracy. Instead, the accuracy achieved is to be
construed as a measure of the stability of the method, which gives an
indication of how well the method is expected to perform under more complex
situations, such as when many channels are involved, or when the range of
the interaction is excessively large, or at very high energies where many
oscillations in the wave functions are present.

In conclusion, all three methods performed well in providing a numerical
solution to the coupled channel test case examined here. This case could
serve as a benchmark calculation for testing additional methods, since the
effective range and the scattering length are also calculated. Further
comparison of various methods under more complex conditions would be
desirable in order to determine the conditions under which a particular one
of the methods would be preferable.

\vspace{0.5in}

Acknowledgments:\ E.T. would like to acknowledge useful discussions with Dr.
Fred Mies, who has implemented the original version of the Gordon method
used by the authors.

\section{Appendix 1. Perturbative treatment of the long-range corrections.}

As the value of $R_{\max }$ (this is the truncation radius beyond which all
potentials are set to zero) is increased, the values of $K_1$ and $K_2$,
defined in Eq. (\ref{DEF}) change slightly because of the non-zero value of
the potential at large distances. In order to extrapolate the $K$ values to $%
R_{\max }=\infty ,$ it is preferable to include the long-range tail of the
potential perturbatively rather than numerically. Showing how this can be
done is the purpose of this appendix. It is of course also possible to
include the long range tails numerically, as has been done in Ref. \cite
{tails} The variation of the $K$'s with large distances, as given in Table
3, can be reproduced numerically by means of perturbation theory, as will
now be shown.

For this section we assume that the radial distance is sufficiently large so
that the coupling potential between the two channels is negligible. Thus,
only the effect of the diagonal potential in each channel $i$ on $K_i$ needs
to be considered. We also continue to assume that the angular momentum
number $\ell $ is zero. Both assumptions can removed by a generalization of
the present discussion.

\subsection{The positive energy channel.}

We will denote two consecutive values of $R_{\max }$ by $T_1$ and $T_{2.},$
respectively, and assume that the numerical calculation extends out to $T_1.$
We denote by $V_1(r)$ the diagonal potential for $r\leq T_1$ and set $%
V_1(r)=0$ for $r>T_1.$ The corresponding wave function is denoted by $\psi
_1 $ which is equal to the wave function $\psi _P$ known numerically.
Because $V_1(r)=0$ for $r>T_1,$ 
\begin{equation}
\psi _1(r)=\sin (kr)+K_1(T_1)\cos (kr).
\end{equation}
Similarly, $V_2(r)$ is equal to the diagonal potential for $r\leq T_2$ and $%
V_2(r)=0$ for $r>T_2.$ The corresponding wave function, to be calculated
perturbatively, is denoted by $\psi _2.$ It obeys the equation $\left(
d^2/dr^2-V_2+k^2\right) \psi _2=0$, which can be written 
\begin{equation}
\left( d^2/dr^2+k^2\right) \psi _2=V^{\prime }\psi _2\;\;r\geq T_1,
\end{equation}
where 
\begin{eqnarray}
V^{\prime }(r) &=&V_2(r)\;\;for\;r\geq T_1 \\
V^{\prime }(r) &=&0\,\,\;\;\;\;\;\;for\;r<T_1.  \nonumber
\end{eqnarray}
Thus, $V^{\prime }$ is the perturbative potential which vanishes outside of
the interval $[T_1,T_2].$ Using for the inverse of the operator $\left(
d^2/dr^2+k^2\right) $ the Green's function integral expression in terms of $%
\sin (kr_{<})\times \cos (kr_{>}),$ one obtains the most general form for $%
\psi _2$ 
\begin{equation}
\psi _2(r)=\alpha \sin (kr)+\beta \cos (kr)+\sin (kr)\Sigma _c(r)+\cos
(kr)\Sigma _s(r)  \label{pert2}
\end{equation}

where 
\begin{equation}
\Sigma _c(r)=-\frac 1k\int_r^{T_2}\cos (kr^{\prime })V^{\prime }(r^{\prime
})\psi _2(r^{\prime })dr^{\prime }
\end{equation}
and 
\begin{equation}
\Sigma _s(r)=-\frac 1k\int_{T_1}^r\sin (kr^{\prime })V^{\prime }(r^{\prime
})\psi _2(r^{\prime })dr^{\prime }.
\end{equation}
The coefficients $\alpha $ and $\beta $ are obtained by matching $\psi _2$
to $\psi _1$ at $r=T_1,$ i.e., setting the two functions and their
derivatives equal. Remembering that $\Sigma _s(T_1)=0,$ that $(1/k)\left(
d\psi _2/dr\right) _{r=T_1}=\alpha s_1+c_1\left( -\beta +\Sigma
_c(T_1)\right) ,$ and that $(1/k)\left( d\psi _1/dr\right)
_{r=T_1}=c_1-K_1(T_1)s_1,$ where $s_1=\sin (kT_1)$ and $c_1=\cos (kT_1),$
one obtains 
\begin{equation}
\alpha =1-\Sigma _a(T_1),\;\;\beta =K_1(T_1).
\end{equation}
Inserting the above result into Eq. (\ref{pert2}), one obtains for $r\geq
T_2 $ the result 
\begin{equation}
\psi _2(r)=\left( 1-\Sigma _c(T_1)\right) \sin (kr)+\left( K_1(T_1)+\Sigma
_s(T_2)\right) \cos (kr).
\end{equation}

The first order perturbation result for the above expressions is obtained by
replacing $\psi _2$ by $\psi _1$ in the integrals for $\Sigma _c$ and $%
\Sigma _s$ in Eqs.above. Denoting the corresponding integrals from $T_1$ to $%
T_2$ by $I_c$ and $I_s,$ respectively 
\begin{equation}
I_c=-\frac 1k\int_{T_1}^{T_2}\cos (kr^{\prime })V^{\prime }(r^{\prime })\psi
_1(r^{\prime })dr^{\prime }
\end{equation}
\begin{equation}
I_s=-\frac 1k\int_{T_1}^{T_2}\sin (kr^{\prime })V^{\prime }(r^{\prime })\psi
_1(r^{\prime })dr^{\prime },
\end{equation}
one finally obtains the first order result for $K_1(T_2)$%
\begin{equation}
K_1^{(1)}(T_2)=\frac{K_1(T_1)+I_s}{1-I_c}\simeq K_1(T_1)+I_s+K_1(T_1)\times
I_c.  \label{pert1}
\end{equation}

This is our final expression. The numerical evaluation was carried out
initially by using MATHEMATICA \cite{W}to evaluate the integrals involving
products of circular functions and $1/r^{6}$. For example, if $T_{1}$ and $%
T_{2}$ are set equal to $500$ and $2000a_{0}$ respectively, and using for $%
C_{6}$ the value stated above, one finds $I_{c}=-8.443\times 10^{-5}$ and $%
I_{s}=-1.6105\times 10^{-5}$. Using for $K_{2}(500)$ the value from Table 3,
one obtains from Eq. (\ref{pert1}) the result 
\[
K_{2}^{(1)}(2000)-K_{2}(500)=1.0265\times 10^{-5}, 
\]
which compares very well with the numerical result from Table 3 
\[
K_{2}(2000)-K_{2}(500)=1.0264\times 10^{-5}. 
\]

The results for the scattering length and the effective range, described in
the text, were calculated by a FORTRAN code which expresses the integrals $%
\int_x^\infty [\sin (t)/t]dt$ and $\int_x^\infty [\cos (t)/t]dt$ in terms of
the functions Ci and Si. The latter were called from the IMSL scientific
library, and generalization to larger powers of $t$ in the denominator (like 
$t^6$ in the present case) were obtained recursively through integrations by
part.

\subsection{The negative energy channel.}

For the negative energy channel a similar perturbative procedure will now be
described. In what follows $\psi _{1}$ and $\psi _{2}$ now refer to the
negative energy channel but otherwise they have the same meaning. The
function $\psi _{1}$ coincides with the numerical solution $\psi _{N}$ for $%
r\leq T_{1},$ while $\psi _{2}$ for $T_{1}\leq r$ is given by 
\begin{equation}
\psi _{2}(r)=\gamma G_{2}(r)-\frac{1}{\kappa }F_{2}(r)%
\int_{r}^{T_{2}}G_{2}V^{\prime }\psi _{2}dr^{\prime }-\frac{1}{\kappa }%
G_{2}(r)\int_{T_{1}}^{r}F_{2}V^{\prime }\psi _{2}dr^{\prime }.  \label{pert3}
\end{equation}
Here $F_{2}(r)$ and $G_{2}(r)$ are equal, respectively to $\exp (-\kappa r)$
and $\sinh (\kappa r).$ By setting $\psi _{2}$ equal to $\psi _{1}$ at $%
r=T_{1}$ one obtains 
\[
\gamma =\frac{1}{G_{2}(T_{1})}\left( \psi _{1}(T_{1})-\frac{1}{\kappa }%
\int_{T_{1}}^{T_{2}}G_{2}(r^{\prime })V^{\prime }(r^{\prime })\psi
_{2}(r^{\prime })dr^{\prime }F_{2}(T_{1})\right) , 
\]
Inserting the above into Eq.(\ref{pert3}) evaluated at $r=T_{2},$ replacing $%
\psi _{2}$ in the integrals in the above expressions by 
\[
\psi _{2}(r)\simeq K_{2}(T_{1})G_{2}(r) 
\]
and dividing the result by $G_{2}(T_{2}),$ one obtains a preliminary value
for $K_{2}(T_{2})$ 
\[
\bar{K}_{2}(T_{2})=K_{2}(T_{1})-\frac{F_{2}(T_{1})}{G_{2}(T_{1})}%
I_{e}+I_{hs} 
\]
where 
\[
I_{e}=-\frac{K_{2}(T_{1})}{\kappa }\int_{T_{1}}^{T_{2}}G_{2}(r^{\prime
})V^{\prime }(r^{\prime })G_{2}(r^{\prime })dr^{\prime } 
\]
\begin{equation}
I_{hs}=-\frac{K_{2}(T_{1})}{\kappa }\int_{T_{1}}^{T_{2}}F_{2}(r^{\prime
})V^{\prime }(r^{\prime })G_{2}(r^{\prime })dr^{\prime }
\end{equation}
The final value of $K_{2}(T_{2})$ is obtained by dividing $\bar{K}%
_{2}(T_{2}) $ by the same normalizing factor $(1-I_{c})$ which was required
to normalize the wave function in channel 1 so that the coefficient in front
of $\sin (kr) $ be equal to $1.$ The final expression is 
\begin{equation}
K_{2}(T_{2})=\frac{\bar{K}_{2}(T_{2})}{(1-I_{c})}\simeq
K_{2}(T_{1})(1+I_{c})-\frac{F_{2}(T_{1})}{G_{2}(T_{1})}I_{e}+I_{hs}
\label{pert4}
\end{equation}

Using for $T_{1}$ and $T_{2}$ the values $500a_{0}$ and $2000a_{0}$
respectively, numerical evaluation of Eq. (\ref{pert4}) gives $%
K_{2}(T_{2})-K_{2}(T_{1})\simeq -5.445\times 10^{-4}$ while the numerical
value obtained from Table 3 gives $\simeq -5.430\times 10^{-4}$\nolinebreak
\

\newpage

\end{document}